\documentclass[12pt,a4paper,twoside,fleqn]{article} 
\usepackage{latexsym} 
\usepackage{emlines} 
\usepackage{epsf} 
\usepackage{feynmf} 
 
\begin{document} 
 
\newcommand{\sumint}{{\textstyle\sum}\hspace{-1.02em}\int} 
\newcommand{\ket}[1]{\left|#1\right>} 
\newcommand{\bra}[1]{\left<#1\right|}

\vspace*{-2cm} 
\begin{center} 
{\Large \bf Trembling cavities 
\vspace{0.1cm} 
 
in the canonical approach} 
 
\end{center} 
\normalsize 
\bigskip 
\begin{center} 
 
\vspace{0.3cm} 
{\large Ralf Sch\"utzhold, G\"unter Plunien and Gerhard Soff} 
  
{\sl Institut f\"ur Theoretische Physik, Technische Universit\"at Dresden,\\ 
Mommsenstr. 13, D-01062 Dresden, Federal Republic of Germany}  
 
\end{center}

\vspace{1.0cm} 
 
\begin{abstract} 
 
We present a canonical formalism facilitating investigations of the dynamical 
Casimir effect 
by means of a response theory approach.   
We consider a massless scalar field confined inside of an arbitaray domain 
$G(t)$, which undergoes 
small displacements for a certain period of time. Under rather general 
conditions a formula for the number of 
created particles per mode is derived.
The pertubative approach reveals the occurance of two generic processes 
contributing to the particle production:
the squeezing of the vacuum by changing the shape and an acceleration effect 
due to motion af the boundaries.  
The method is applied to the configuration of moving mirror(s). 
Some properties as well as  
the relation to local Green function methods are discussed.  
 
\end{abstract} 
 
\bigskip 
 
PACS-numbers:  12.20; 42.50; 03.70.+k; 42.65.Vh\\  
\vspace{0.1cm}

Vibrating boundary;

\section{Introduction} 
 
After the discovery of the static Casimir effect \cite{s1} 
(see, e.g. Refs. \cite{s2}, \cite{s3} and \cite{s4} for a review),  
the possibility 
of creating particles out of the vacuum by moving one of the mirrors 
(see, e.g. \cite{p1}-\cite{p0}) 
or both plates (see, e.g. \cite{b1}) has been analyzed. Other authors 
calculated the radiation from 
one single mirror (see \cite{m1}-\cite{m5}) and the backreaction on 
perfectly or nonperfectly 
conducting boundaries (see, e.g. \cite{f1}-\cite{f0}).  
Many intresting studies also have been devoted to the analysis of quantum 
vacuum radiation induced by moving
dielectrics \cite{medium}. 
We discuss a general Hamiltonian formalism for an arbitary domain $G(t)$ with 
Dirichlet boundary conditions  
that experiences small changes during a time interval $(0,T)$:  
 
\begin{equation} 
G(t<0)=G(t>T)=G_0. 
\end{equation}

In  our derivations we imply that the time-dependent disturbances 
$\Delta G(t)=G(t) \ominus G_0$ of the 
boundary can be considered as small with respect to some parameter 
$\varepsilon$, i.e.  
$\Delta G = {\cal O}(\varepsilon)$. Having introduced a proper definition of 
particles together with 
a vacuum state, we calculate the number of produced particles within the 
framework of response theory. 
The result will be applied to the special case of moving mirror(s) and the 
relation to results obtained 
by means of the adiabatic approach or local Green function methods will be 
indicated.

\section{Canonical formulation} 
 
\subsection{Equations of motion} 
 
We consider a non-interacting real massless scalar (Klein-Gordon) field  
in Minkowski-space-time with the Dirichlet-boundary-conditions: 
$\Phi=0$ at $\partial G(t)\;$ and $\;\Box\,\Phi=0$ in $G(t)$. 
In the following we shall not quantize the degrees of freedom assigned to the 
motion of the 
boundary.  
Focussing on the scalar field sector only, we consider the Lagrangian
($\hbar=c=1$ throughout) 
 
\begin{equation} 
{\cal L}=\frac{1}{2}\partial_\mu\Phi\partial^\mu\Phi 
\label{eq:1} 
\end{equation} 
 
Expanding $\Phi$ in terms of eigenfunctions $f_\alpha(\vec r,t)$ that fullfil  
$f_\alpha=0$ at the boundary $\partial G(t)$, i.e.,
 
\begin{equation}  
\label{ex} 
\Phi(\vec r,t)=\sumint\limits_\alpha\,q_\alpha(t)\,f_\alpha(\vec r,t)   
\end{equation} 
 
and calculating the Lagrangian by making use of the properties Eqs. 
(\ref{g2})-(\ref{g4}) given in the appendix, we arrive at 
 
\begin{equation}  
\label{l} 
L=\int\limits_{G(t)}\,dV\,{\cal L}= 
\frac{1}{2}\dot{q}^2_\alpha-\frac{1}{2}\Omega^2_\alpha(t)q^2_\alpha 
+q_\alpha {\cal M}_{\alpha\beta}(t)\dot{q}_\beta 
+\frac{1}{2}q_\alpha{\cal M_{\alpha\gamma}}(t){\cal M_{\beta\gamma}}(t)q_\beta 
\end{equation} 
 
together with the eigenvalue equation 
 
\begin{equation}  
\nabla^2\,f_\alpha(\vec r,t)=-\Omega^2_\alpha(t)\,f_\alpha(\vec r,t) 
\end{equation} 
 
and the coupling matrix  
 
\begin{equation}  
{\cal M}_{\alpha\beta}(t)=\int\limits_{G(t)}\,dV 
\frac{\partial f_\alpha(\vec r,t)}{\partial t}\,f_\beta(\vec r,t) \quad . 
\end{equation} 
 
In view of the orthonormality of the $f_\alpha$ 
(see Eq. (\ref{g2}) in the appendix; $dG=dV$; $d\dot G=\vec v\,d\vec A\,$) 
and the required boundary-conditions the ${\cal M}_{\alpha\beta}$ turn out to 
be antisymmetric: 
 
\begin{eqnarray}  
\label{asy}\nonumber 
{\cal M}_{\alpha\beta}(t)+{\cal M}_{\beta\alpha}(t) 
&=&\int\limits_{G(t)}dG 
\frac{\partial}{\partial t}[f_\alpha(\vec r,t)f_\beta(\vec r,t)]\\ 
&=&\frac{d}{dt}\delta(\alpha,\beta) 
-\int\limits_{\partial G(t)}d\dot G\,f_\alpha(\vec r,t)f_\beta(\vec r,t)=0 
\end{eqnarray}

In Eq. (\ref{l}) and in most of the following formulae we drop the summation- 
and integration signs and declare that one has to sum over all multi-indices 
like $\alpha$,$\beta$ etc. that do not occur on both sides of the equation. 
Introducing the canonical conjugate momenta  
 
\begin{equation}  
p_\alpha=\frac{\partial L}{\partial\dot q_\alpha}= 
\dot q_\alpha+q_\beta{\cal M}_{\beta\alpha}(t) \quad , 
\end{equation} 
 
the Hamiltonian takes on the form: 
 
\begin{equation}  
H(t)=\frac{1}{2}p^2_\alpha+\frac{1}{2}\Omega^2_\alpha(t)q^2_\alpha+ 
p_\alpha{\cal M}_{\alpha\beta}(t)q_\beta \quad . 
\label{eq:9} 
\end{equation} 
 
There are two effects which could lead to an unstable vacuum:  
The nonstationary eigenfrequencies $\Omega_\alpha(t)$ due to a dynamical  
change of the shape of the domain $G(t)$ -- we shall refere to this  
effect as ``squeezing'' of the vacuum -- 
and the additional $q_\alpha$-$p_\beta$-coupling ${\cal M}_{\alpha\beta}$, 
indicating the motion of the boundaries -- the ``acceleration''-effect. 
The total energy of the $\Phi$-field is given as the integral over the 
time-dependent 
domain $G(t)$ of the energy density $T_{00}$ 
 
\begin{equation}  
E(t)=\int\limits_{G(t)}\,dV\,T_{00}= 
\frac{1}{2}p^2_\alpha+\frac{1}{2}\Omega^2_\alpha(t)q^2_\alpha 
\end{equation} 
 
of the minimal coupled energy-momentum tensor: 
 
\begin{equation}  
T_{\mu\nu}=\partial_\mu\Phi\partial_\nu\Phi 
-\frac{1}{2}g_{\mu\nu}\partial_\rho\Phi\partial^\rho\Phi \quad . 
\end{equation} 
 
Comparison with Eq. (\ref{eq:9}) reveals that 
 
\begin{equation} 
H(t)=E(t)+W(t) 
\end{equation} 
 
holds. 
The time-dependent transformation $\Phi (\vec r,t) \rightarrow q_\alpha(t)$ 
results in 
the difference between the Hamiltonian, describing the time-evolution of the 
$q_\alpha$,  
and that of $\Phi$ (which is equal to the field energy):   
 
\begin{equation}  
H[q_\alpha,p_\beta,t]\neq  
H[\Phi,\Pi,t]= 
E[\Phi,\Pi,t]= 
E[q_\alpha,p_\beta,t] \quad . 
\end{equation}

\subsection{Quantization} 
 
Now we perform the usual canonical quantization, assuming the following set\\ 
of equal-time-commutation-relations: 
 
\begin{equation}  
[\hat q_\alpha(t),\hat q_\beta(t)]= 
[\hat p_\alpha(t),\hat p_\beta(t)]=0 
\quad , 
\end{equation} 
 
\begin{equation}  
[\hat q_\alpha(t),\hat p_\beta(t)]=i\delta(\alpha,\beta) 
\quad . 
\end{equation} 
 
Note that, because of (\ref{ex}) and (\ref{g4}) together with the decomposition 
\begin{equation} 
\label{imp} 
\Pi(\vec r,t)=\dot\Phi(\vec r,t)= 
p_\alpha(t)\,f_\alpha(\vec r,t) 
\end{equation} 
 
of the conjugate momenta these commutation relations are consistent with those  
between the fields: 
 
\begin{equation}  
[\hat\Phi(\vec r,t),\hat\Phi(\vec r\,',t)]= 
[\hat \Pi(\vec r,t),\hat \Pi(\vec r\,',t)]=0 
\end{equation} 
 
and 
 
\begin{equation}  
\label{delta} 
[\hat\Phi(\vec r,t),\hat\Pi(\vec r\,',t)]= 
i\delta(\vec r-\vec r\,') \quad , 
\end{equation} 
 
which are valid inside the domain $G(t)$.
Equation (\ref{imp}) and therefore Eq. (\ref{delta}) are not pointwise 
equalities (think of $\partial G$), 
they have to be read as identities of $L_2(G)$-distributions.

\section{Vacuum-definition} 
 
\subsection{Interaction-representation} 
   
For performing the pertubation-theory we shall adopt the interaction 
representation. 
Accordingly, the time-evolution of the operators will be governed by the 
undisturbed energy  
operator $\hat E_0=\hat E(t<0)=\hat E(t>T)$ defined via  
 
\begin{equation}  
\hat H(t)=\hat E(t)+\hat W(t)= 
\hat E_0 +\Delta\hat E(t)+\hat W(t)= 
\hat E_0+\hat H_1(t)=\hat H_0+\hat H_1(t) \quad . 
\end{equation} 
 
Time-dependent operators $\hat{A}(t)$ obey the equation of motion 
 
\begin{equation}  
\frac{d\hat A}{dt}= 
i[\hat E_0 , \hat A]+\frac{\partial\hat A}{\partial t} \quad , 
\end{equation} 
 
while the dynamics of any given quantum state $\ket{\psi}$ is described by  
 
\begin{equation}  
\label{psi} 
\frac{d}{dt}\ket\psi=-i\hat H_1(t)\ket\psi  \quad , 
\end{equation} 
 
where the interaction Hamiltonian $\hat{H}_1$ is specified in Eqs. 
(\ref{h1})--(\ref{dw}) 
below. 
 
\subsection{The number operator} 
 
A proper set of particle creation and annihilation operators should be 
introduced 
with respect to the unperturbed Hamiltonian  
$\hat{H}_0 = \hat{H}(t<0)=\hat{H}(t>T)=\hat{E}_0$ according to  
 
\begin{equation}  
\hat a_\alpha(t)=(2\Omega^0_\alpha)^{-1/2} 
(\Omega^0_\alpha\hat q_\alpha(t) + i\hat p_\alpha(t)) \quad , 
\end{equation} 
 
together with the static frequencies $\Omega^0_\alpha=\Omega_\alpha(t<0)=
\Omega_\alpha(t>T)$.\\ 
They obey the equation of motion: 
 
\begin{eqnarray} 
\frac{d \hat{a}_\alpha }{dt}= i [\hat{E}_0,\hat{a}_\alpha]=
-i\Omega^0_\alpha\hat a_\alpha 
\end{eqnarray} 
 
and the commutation relation 
 
\begin{equation}  
[\hat a_\alpha(t),\hat a^{+}_\beta(t)]=\delta(\alpha,\beta)\quad . 
\end{equation} 
 
These particle creation and annihilation operators diagonalize the unperturbed 
Hamiltionian, i.e.,  
expressed in terms of the corresponding number operator 
 
\begin{equation} 
\hat N_\alpha=\hat a^{+}_\alpha(t)\hat a_\alpha(t) 
\end{equation} 
 
it thus takes the form 
 
\begin{equation}  
\hat E_0=\Omega^0_\alpha\left(\hat N_\alpha +\frac{1}{2}\right)\quad . 
\end{equation} 
 
Evidently we define the vacuum $\ket{0}$ as the ground state of 
$\hat E_0$: 
 
\begin{equation}  
\forall\alpha\;:\; \hat a_\alpha\ket{0}=0 \quad . 
\end{equation} 
 
So $\hat N_\alpha$ counts the physical relevant particles in the mode $\alpha$ 
before and after the motion 
of the boundaries (due to $\ket{0,in}=\ket{0,out}$), a particle definition 
during the movement is not  
so easy to obtain (see section 5 and 6).  
 
\section{Particle creation} 
 
\subsection{Response theory} 
 
Now we investigate the change of the state vector $\ket\psi$, which satisfies 
the initial 
condition  
 
\begin{equation}  
\ket{\psi(t<0)}=\ket{0}\quad , 
\end{equation} 
 
due to a small time-bounded but otherwise arbitrary motion of the boundary by 
computing the 
number of created particles per mode $\alpha$ to first non-vanishing order 
perturbation theory.  
Equation (\ref{psi}) can be formally integrated by means of the time ordering 
operator ${\cal T}$ 
 
\begin{eqnarray}  
\ket{\psi(T)} 
&=& {\cal{T}}\left[\exp\left\{-i\int\limits^T_0\,dt\,\hat H_1(t)\right\}\right]
\ket{0}\\ 
\nonumber 
&=& \sum_{n=0}^\infty \,\frac{(-i)^n}{n!}\, 
\int_0^T dt_n \cdots \int_0^T dt_1\, {\cal T}\,  
\left[\hat{H}_1(t_n)\cdots \hat{H}_1(t_1)\right] \ket{0} \quad . 
\end{eqnarray} 
 
Assuming small perturbations we shall keep only the lowest-order terms of the 
expansion 
above. The time-evolved vacuum state reads 
 
\begin{equation}  
\ket{\psi(T)}= 
\left[1-i\int\limits^T_0\,dt\;\hat H_1(t)\right]\ket{0} 
+{\cal O}(\hat H^2_1)\quad . 
\end{equation} 
 
In view of the property 
 
\begin{equation}  
\forall\alpha\,\,\,\hat N_\alpha\ket{0}=0  
\end{equation} 
 
the number operator has no linear response; the first non-vanishing order is 
quadratic. 
But other operators $\hat A$ with $\hat A \ket{0}\neq0$ such as e.g. the 
components of the energy-momentum tensor $\hat T_{\mu\nu}$, possess a linear 
response: 
 
\begin{equation} 
\bra{\psi(T)}\hat A\ket{\psi(T)}= 
\bra{0}\hat A\ket{0}+ 
i\int\limits^T_0\,dt\bra{0}[\hat H_1(t),\hat A]\ket{0}+ 
{\cal O}(\hat H^2_1) \quad . 
\end{equation} 
 
To be complete we note here the general expression for the quadratic response: 
 
\begin{eqnarray}\nonumber 
\bra{\psi(T)}\hat A\ket{\psi(T)}&=& 
\bra{0}\hat A\ket{0}+ 
i\int\limits^T_0\,dt\bra{0}[\hat H_1(t),\hat A]\ket{0}\\ 
& &\hspace{-3cm}+\int\limits^T_0\,dt\int\limits^T_0\,dt'\,\left( 
\bra{0}\hat H_1(t)\hat A\hat H_1(t')\ket{0}- 
\frac{1}{2}\bra{0}\{{\cal T}[\hat H_1(t)\hat H_1(t')],\hat A\}\ket{0}\right) 
+{\cal O}(\hat H^3_1) 
\quad . 
\end{eqnarray} 
 
We are now in the position to calculate the number of particles $N_\alpha$ 
created in the 
specific mode $\alpha$ after the time duration $T$, when the boundaries are 
again at rest.  
We have to evaluate the matrix element 
 
\begin{equation}   
\bra{\psi(T)}\hat N_\alpha\ket{\psi(T)}= 
\int\limits^T_0\,dt\int\limits^T_0\,dt'\, 
\bra{0}\hat H_1(t)\hat N_\alpha\hat H_1(t')\ket{0} 
+{\cal O}(\hat H^3_1)= 
N_\alpha+{\cal O}(\hat H^3_1) 
\end{equation} 
 
with the interaction Hamiltonian 
 
\begin{equation}  
\label{h1} 
\hat H_1(t)=\Delta \hat E(t) + \hat W(t)\quad , 
\end{equation} 
 
where 
 
\begin{equation}  
\label{de} 
\Delta\hat E(t)= 
\frac{1}{2}\hat q^2_\alpha(t)\;\Delta\Omega^2_\alpha(t) 
\end{equation} 
 
and 
 
\begin{equation} 
\label{dw}  
\hat W(t)=\hat q_\alpha(t) {\cal M_{\alpha\beta}}(t)\hat p_\beta(t) 
\quad . 
\end{equation} 
 
According to Eq. (\ref{asy}) $\;{\cal M_{\alpha\alpha}}(t)=0$ holds and thus 
the cross terms 
vanish, i.e.  
 
\begin{equation}  
\forall\alpha\,\,\, 
\bra{0}\hat W(t) \hat N_\alpha \Delta \hat E(t') \ket{0}=0\quad . 
\end{equation} 
 
As a consequence, one obtains 
 
\begin{equation}  
N_\alpha=  
\int\limits^T_0\,dt\int\limits^T_0\,dt'\, 
(\bra{0}\Delta\hat E(t)\hat N_\alpha\Delta\hat E(t')\ket{0}+ 
\bra{0}\hat{W}(t)\hat{N}_\alpha \hat{W}(t')\ket{0}) 
\quad. 
\end{equation} 
 
Up to quadratic order the sqeezing- (first term) and the acceleration-effect 
(second term) decouple, 
so that:  $N_\alpha=N^S_\alpha+N^A_\alpha$. 
With the aid of Eqs. (\ref{de}) and (\ref{dw}) we get: 
 
\begin{eqnarray}  
N_\alpha &=&  
\int\limits^T_0\,dt\int\limits^T_0\,dt'\, 
\frac{1}{4}\Delta\Omega^2_\beta(t)\,\Delta\Omega^2_\gamma(t')\, 
\bra{0}\hat q^2_\beta(t)\hat N_\alpha\hat q^2_\gamma(t')\ket{0}  
\nonumber \\ 
&+& 
\int\limits^T_0\,dt\int\limits^T_0\,dt'\, 
{\cal M}_{\kappa\lambda}(t){\cal M}_{\sigma\tau}(t)\, 
\bra{0}\hat q_\kappa(t)\hat p_\lambda(t) 
\hat N_\alpha 
\hat q_\sigma(t')\hat p_\tau(t')\ket{0} \quad. 
\end{eqnarray} 
 
Evaluation of the expectation values by utilizing the equation of motion 
(in the $\hat E_0$-dynamic) leads to 
 
\begin{eqnarray} 
\label{zerg}  
N_\alpha &=&  
\int\limits^T_0\,dt\int\limits^T_0\,dt'\, 
\frac{1}{4(\Omega^0_\alpha)^2} 
\Delta\Omega^2_\alpha(t)\,\Delta\Omega^2_\alpha(t')\, 
\exp(2i\Omega^0_\alpha[t'-t])\nonumber\\ 
&+& 
\int\limits^T_0\,dt\int\limits^T_0\,dt'\, 
{\cal S}_{\alpha\beta}(t){\cal S}_{\alpha\beta}(t')\, 
\exp(i[\Omega^0_\alpha+\Omega^0_\beta][t'-t]) 
\end{eqnarray} 
 
with the symmetric matrix: 
 
\begin{equation}  
\label{Sab} 
{\cal S}_{\alpha\beta}(t)={\cal S}_{\beta\alpha}(t)= 
\frac{1}{2}{\cal M}_{\alpha\beta}(t)\, 
\left(  
\sqrt{\frac{\Omega^0_\beta}{\Omega^0_\alpha}}-
\sqrt{\frac{\Omega^0_\alpha}{\Omega^0_\beta}}
\;\right) \quad . 
\end{equation} 
 
A more compact form of Eq. (\ref{zerg}) is derived by using the 
Fourier-transformation ${\cal F}$\\  
with  $\varphi(t)\rightarrow\widetilde\varphi(\omega)=
[{\cal F}\varphi](\omega)$  :  
 
\begin{eqnarray}  
\label{erg} 
N_\alpha= 
\frac{1}{4(\Omega^0_\alpha)^2} 
|\widetilde{\Delta\Omega^2_\alpha}(2\Omega^0_\alpha)|^2\,+\, 
|\widetilde{{\cal S}_{\alpha\beta}}(\Omega^0_\alpha+\Omega^0_\beta)|^2 = 
N^S_\alpha + N^A_\alpha \quad . 
\end{eqnarray} 
 
This spectral representation above provides the main result of the perturbative
approach:  
The number of particles created in the mode $\alpha$ (with an energy 
$\Omega^0_\alpha$) decomposes 
into a squeezing and an accelleration contribution. The occurance of these two 
distinct contributions  
reflects the two basic degrees of freedom that characterize dynamical changes 
of the boundaries: 
deformations in shape and motion of the boundaries. The total number of 
particles 
produced after the time interval $(0,T)$ is obtained by summing/integrating 
over all modes $\alpha$: 
 
\begin{eqnarray} 
N = \sumint\limits_\alpha \, N_\alpha\quad . 
\end{eqnarray}  
 
However, this quantity is in general ill-defined and requires an appropriate 
regularization.  
This may be most easily achieved by introducing explicitly a frequency cut-off 
which simulates 
imperfect conducting boundary conditions. Whether a cut-off-independent 
contribution remains after 
its removal will depend on the particular configuration of boundaries under 
consideration.     
We should also note, that for ''well behaving'' time dependencies the spectral 
form (\ref{erg}) 
may provide sufficient convergence of the summation/integration in the limit 
of perfect conductors. 
 
\subsection{Discussion} 
 
Formula (\ref{erg}) gives the produced number of particles in a given mode 
$\alpha$ in lowest order 
pertubation-theory in $\hat H_1$. If we want to analyze a particular situation 
within the 
perturbative approach we need to specify the very general quantity $\hat H_1$. 
Furthermore, we have to introduce the magnitude of the displacement of the 
boundaries $\varepsilon$. 
By inspection, we see that all quantities entering $\hat{H}_1$ are at least of 
order $\varepsilon$,  
i.e.,  $\Delta\Omega^2_\alpha={\cal O}(\varepsilon)$ and  
${\cal M}_{\alpha\beta}={\cal O}(\varepsilon)$ (and $\Delta f_\alpha=
{\cal O}(\varepsilon)$ as well) 
reveals that $\hat H_1(t)={\cal O}(\varepsilon)$. This already indicates that 
both, the squeezing- and the acceleration-effect are of the same order of 
magnitude. 
As a matter of fact, the adiabatic approximation applied frequently in studies 
cannot be sufficient in general. 
In the adiabatic approach the acceleration term (second term) is not obtained.  
Consequently, the dynamics reduces to a set of decoupled ordinary differential 
equations of the form: 
 
\begin{equation} 
\ddot x_\alpha(t)\;+\;\Omega^2_\alpha(t)\,x_\alpha(t)\,=0\quad. 
\end{equation} 
 
They can be solved by means of a scattering-theory approach 
(see e.g. \cite{p5}). The adiabatic approach 
thus mainly accounts for the squeezing contribution. 
We like to stress that there are indeed situations, where $N^S_\alpha$ is much 
smaller than  
$N^A_\alpha$. Let $\Omega_\alpha(t)$ have a functional form corresponding to a 
reflectionless potential, for example: 
 
\begin{equation} 
\Delta\Omega^2_\alpha(t)=\frac{2\nu^2}{\cosh^2(\nu t)} 
\end{equation} 
 
one obtains $N^S_\alpha=0$. Or, consider e.g. harmonic oszillations of the 
boundaries with the frequency  
$\omega_0$ over a long period $T$  
with $T\omega_0\gg 1$. In that case  only particles with $\Omega_\alpha=
\frac{1}{2}\omega_0$ are produced by the  
sqeezing effect, but the acceleration effect also leads to the production of 
particles with other frequencies.\\ 
This example already demonstrates the high-resonant character of the squeezing 
term, while the accereration term does not have this 
property due to the summation/integration over the modes. As mentioned above, 
another feature of the Fourier transformation leads to the result, that all 
expressions like total number  
of particles, total energy, etc. are convergent after summation/integration 
over all modes, if (and only if) the  
time-dependend function of the displacement of the boundaries is smooth enough.

\section{Squeezing effects} 
 
To investigate the pure squeezing effect it is convenient to neglect any 
explicit time-dependence of the system. 
In the following considerations we restrict to situations, where the 
time-dependence enters only implicitly via 
some global length parameter $\chi$ describing the changes of the shape of the 
boundary $\partial G$. 
In general $\chi$ could abbreviate a set of suitable parameters characterizing 
the dynamics of the shape. 
While the pure accelleration effect can appear separately e.g. as a rigid 
motion of the boundary as a whole, 
there seem to be no possible $G(t)$ that will lead in a pure sqeezing effect 
only, since 
there is no shape-changing without inducing motions of the boundaries. 
Accordingly, the $\Omega_\alpha$ become functions of $\chi$ and therefore also
the creation- and  
annihilation-operators as well:  
 
\begin{equation} 
 \hat a_\alpha(\chi)=(2\Omega_\alpha(\chi))^{-1/2} 
(\Omega_\alpha(\chi)\hat q_\alpha + i\hat p_\alpha) 
\quad . 
\end{equation} 
 
The same holds for the total energy of the field: 
 
\begin{equation}  
\hat E(\chi)=\Omega_\alpha(\chi)\left(\hat N_\alpha(\chi) +\frac{1}{2}\right)=
\hat E_N+E_Z \quad ,
\end{equation} 

where $E_Z$ denotes the zero-point energy and $\hat E_N$ that of the contained 
particles.\\ 
Equivalently,  
 
\begin{equation} 
\forall\alpha\;:\; \hat a_\alpha(\chi)\ket{0(\chi)}=0   
\end{equation} 
 
for the vacuum as the ground state of $\hat E(\chi)$. 
Now we investigate the potential arising from the constrained vacuum of the 
quantized scalar field, 
which may be expanded around a fixed but arbitrary (shape) configuration $\chi 
=\chi_0 + \Delta\chi$:  
 
\begin{equation} 
V(\chi)= 
\bra{0(\chi_0)}\hat E(\chi)\ket{0(\chi_0)}= 
V(\chi_0)+ 
\left(\frac{\partial V}{\partial \chi}\right)_{\chi_0} 
\!\!\!\!\Delta\chi+ 
\frac{1}{2}\left(\frac{\partial ^2V}{\partial \chi^2}\right)_{\chi_0} 
\!\!\!\!\Delta\chi^2+ 
{\cal O}(\Delta\chi^3)\quad . 
\end{equation} 
 
Formally we can introduce a force via 
 
\begin{equation}  
F(\chi_0)=  
\left(\frac{\partial V}{\partial \chi}\right)_{\chi_0}=
\left(\frac{\partial E_Z}{\partial \chi}\right)_{\chi_0}=
\left(\frac{\partial\Omega_\alpha}{\partial \chi}\right)_{\chi_0}
\quad , 
\end{equation} 
 
which should lead to the well-known Casimir force after appropriate 
regularizations have  
been performed \cite{eliz}.  
However, the second-derivative term   
 
\begin{equation}  
\left(\frac{\partial^2 V}{\partial \chi^2}\right)_{\chi_0}=
\left(\frac{\partial^2 E_Z}{\partial \chi^2}\right)_{\chi_0}+
\bra{0(\chi_0)}\left(\frac{\partial^2\hat E_N}{\partial\chi^2}\right)_{\chi_0}
\ket{0(\chi_0)}
\end{equation} 
 
with

\begin{eqnarray}
\label{erzeug}
\bra{0(\chi_0)}\left(\frac{\partial^2\hat E_N}{\partial\chi^2}\right)_{\chi_0}
\ket{0(\chi_0)}&=&
\Omega_\alpha\bra{0(\chi_0)}\left(\frac{\partial^2\hat N_\alpha}{\partial\chi^2}
\right)_{\chi_0}\ket{0(\chi_0)}\\
\nonumber
&=&
2\Omega_\alpha\bra{0(\chi_0)}  
\left(\frac{\partial\hat a^+_\alpha}{\partial\chi}\right)_{\chi_0}  
\left(\frac{\partial\hat a_\alpha}{\partial\chi}\right)_{\chi_0}
\ket{0(\chi_0)}
\end{eqnarray}

and

\begin{equation}
\frac{\partial\hat a_\alpha}{\partial\chi}=\frac{1}{2\Omega_\alpha}\,
\frac{\partial\Omega_\alpha}{\partial\chi}\;\hat a^+_\alpha
\end{equation}

gives rise to an additional parabolic potential due to the fact that    
 
\begin{equation}  
\bra{0(\chi)}\frac{\partial^{2}\hat E}{\partial \chi^2}\ket{0(\chi)}\neq 
\frac{\partial^2}{\partial \chi^2}\bra{0(\chi)}\hat E(\chi)\ket{0(\chi)} 
\quad . 
\end{equation} 
 
As a consequence, after the summation/integration over all modes $\alpha$ is 
performed the additional 
parabolic potential has an infinite strength 
(easy to verify e.g. for the moving mirror example; in contrast to the 
parabolic potential the Casimir force
turns out to be convergent because its divergent parts cancel, if both sides 
of the mirror are taken into account) 
which counteracts to any displacements of the boundary.  
This leads to the conclusion that for an empty closed system of perfectly 
conducting boundaries at  
zero temperature it would be impossible to observe the static Casimir effect 
via measurents of the 
Casimir forces exerted on the boundary.  
As we can see in (\ref{erzeug}), any finite change of $\chi$ would lead to an 
infinite amount of produced particles 
so that the backreaction would compensate the Casimir force even after an 
infinitesimal displacement.

\section{Velocity effects} 
 
The pure acceleration effect can be studied by chosing $G(t)=\vec\eta(t)+G_0$. 
(That means a time dependent translation of the ``rigid'' domain $G_0$, another 
possibility could be a rotation.) 
In view of 
 
\begin{equation} 
\hat W(t)\ket{0}=\frac{i}{2} {\cal S}_{\alpha\beta}(t) 
\hat a^{+}_\alpha\hat a^{+}_\beta\ket{0}\neq 0  
\end{equation} 
 
the ground state of $\hat E_0$ is not stable under the time-evolution of  
$\hat H$, even for constant velocities. Therefore, for nonvanishing  
$\dot{\vec \eta}$, 
the diagonalization of $\hat E_0$ does not yield a proper definition of 
creation/annihilation operators 
$\hat{a}^+_\alpha / \hat{a}_\alpha$ that describe physical 
(i.e. Lorentz-invariant) particles. Only  
in the frame where the boundaries are (globally) at rest a reasonable 
definition of particles by  
diagonalization of $\hat E_0$ is possible. One should notice that:  
 
\begin{equation} 
\hat H(t)=\hat E_0+\hat W(t)= 
\int\limits_{G(t)}\,dV\,\left(\,\frac{1}{2}\, 
[\,\hat\Pi^2+(\nabla\hat\Phi)^2\,]\,-\, 
\hat\Pi\,(\,\dot{\vec\eta}\,\nabla\,)\,\hat\Phi \right)  
\end{equation} 
 
and therefore: 
 
\begin{equation} 
\gamma\hat H= 
\int\limits_{G(t)}dV 
\left(\hat T_{\mu\nu}\,\Lambda^\nu_{\;\rho}(\dot\eta)\right)_{0\,0}= 
\int\limits_{G(t)}dV\, 
\hat T_{0\nu}\,\Lambda^\nu_{\; 0}(\dot\eta)=
\int\limits_{G(t)}d\Sigma^\mu_{G}\;\hat T_{\mu\nu}u^\nu 
\quad , 
\end{equation} 
 
where $\Lambda=\Lambda^\nu_{\;\rho}(\dot\eta)$ denotes the Lorentz 
transformation
with the four-velocity  
$u_\mu=\gamma(1,\dot{\vec\eta}\,)=(1,0,0,\dot\eta)/\sqrt{1-\dot\eta^2}$ of the 
boundary.\\ 
Also for $\gamma\approx 1$ the Hamiltonian as defined above does not coincide 
with the energy operator : 
 
\begin{equation} 
\hat H\neq\hat E_0\:\!'
\,=\, 
\int\limits_{G_0}dV' 
\left(\,\hat T\,'\,\right)_{0\,0} 
\,=\, 
\int\limits_{G_0}dV' 
\left(\,\Lambda^+\,\hat T\,\Lambda\,\right)_{0\,0} 
\,=\, 
\int\limits_{G_0}dV'\, 
u^\mu\,\hat T_{\mu\nu}\,u^\nu  
\end{equation} 
 
introduced by a co-moving observer. 
This is an indication of the fact, that $\hat H(t)$ describes the dynamics for 
the observer time $t$,  
and not for the time $t'$ of the co-moving frame. 
 
\section{Moving-mirrors-configuration} 
 
Now we are going to apply the formalism derived above to the special case of 
two parallel mirrors  
placed at $z=\eta(t)$ and $z=\eta(t)+l(t)$ or one single mirror located at 
$z=\eta(t)$.\\ 
(We assume $\eta(t<0)=\eta(t>T)=0$ and $l(t<0)=l(t>T)=l_0 $.)\\   
With $I(t)=(\eta(t);l(t)+\eta(t))$ it follows (see appendix): 
 
\begin{equation}  
\Delta\Omega^2_\alpha(t)= 
\frac{n^2\pi^2}{l^2(t)}-\frac{n^2\pi^2}{l^2_0}= 
\frac{n^2\pi^2}{l^2_0}\xi(t)= 
(\Omega^\|_\alpha)^2_0\,\xi(t) 
\end{equation} 
 
And then the squeezing term has the simple form: 
 
\begin{equation}  
N^S_\alpha= 
\frac{(\Omega^\|_\alpha)^4_0}{4(\Omega^0_\alpha)^2} 
|\widetilde{\xi}(2\Omega^0_\alpha)|^2 
\quad . 
\end{equation} 
 
For the acceleration term one does not obtain such a simple expression,  
an explicit form follows from (\ref{a14}), (\ref{a15}), (\ref{a16}) and 
(\ref{Sab}), inserted in (\ref{zerg}) or (\ref{erg}).   
With $G(t)=I(t)=(\eta(t);\infty)$ the squeezing term vanishes 
(the same as in the case $l=$const) 
and the acceleration term yields after some simplifications [see (\ref{a17}) 
and (\ref{end})]: 
 
\begin{equation} 
\label{movmir} 
N_\alpha= 
\frac{1}{\pi^2}\,\int\limits^\infty_0\, 
d\Omega^0_\beta\;\;\Omega^0_\alpha\,\Omega^0_\beta\, 
|\widetilde{\eta}(\Omega^0_\alpha+\Omega^0_\beta)|^2 
\quad . 
\end{equation} 
 
(\ref{movmir}) enables us to calculate also the total energy radiated by a 
mirror\\ 
into the 1+1-dimensional $G(t)$: 
 
\begin{equation} 
\label{rad} 
E=\Omega_\alpha^0 N^A_\alpha+{\cal O}(\varepsilon^3) 
 =\frac{1}{12\pi}\int dt\;\ddot\eta^2(t)\;+{\cal O}(\eta^3) 
\quad . 
\end{equation}  
 
\subsection{Mechanical properties} 
 
The force exerted upon each mirror can be calculated by computing the divergence
of the 
symmetrized energy-momentum tensor:  
 
\begin{equation}  
\label{force1} 
\hat f_\nu= 
\partial^\mu \hat T_{\mu\nu}= 
\left \{  ( \partial_\nu\hat\Phi)\Box\hat\Phi    \right \} \quad . 
\end{equation} 
 
$\Box\Phi=0$ is valid only in $G(t)$ but it is not at $\partial G(t)$. E.g., 
a mirror placed at 
$z=\eta(t)$ with $\eta(t=0)=0$ and $\dot\eta(t=0)=0$ induces the following 
source term 
(this can be verified by means of Fourier analysis): 
 
\begin{equation}  
\Box\hat\Phi= 
(\vec n\,\nabla\hat\Phi)\,\delta(z)\quad , 
\end{equation} 
 
where $\vec n$ denotes the normal to the plane mirror. Therefore: 
 
\begin{equation}  
\label{force2}
\hat f_\nu= 
\partial^\mu \hat T_{\mu\nu}= 
\left\{ ( \partial_\nu\hat\Phi ) \vec n\,\nabla\hat\Phi\right\}\delta(z) 
\quad .
\end{equation} 
 
Expressions (\ref{force1})--(\ref{force2}) describe  
the force acting on the mirror at only one side, for a complete examination  
(and for renormalization) it is necessary to take both sides into account.\\  
Taking the vacuum expectation value, we obtain the mechanical force density 
 
\begin{equation}  
\vec f= 
\vec n\bra{0}(\vec n\,\nabla\hat\Phi)^2\ket{0}\delta(z) 
\quad . 
\end{equation} 
 
The corresponding force is obtained after integration over space. 
So a mirror at $t=0$ experiences only the static Casimir force, also for 
non-vanishing $\ddot\eta$ and $\stackrel{...}{\eta}$. 
Other forces (e.g. $\sim \stackrel{...}{\eta}$, see \cite{m1}-\cite{m4} and 
\cite{f1}-\cite{f0}) do not occur at $t=0$  
but possibly at later times, when the state vector describing the system has 
changed: $\ket{\psi}\neq\ket{0}$. 
 
\section{Remaining questions} 
 
Some modifications of the formalism presented so far become necessary, if we 
turn to the electromagnetic field  
(polarizations, gauge, etc.) or if Neumann boundary conditions would be 
required. However, the 
general structure of the formalism and the results presented remain very much 
the same. 
Another possible  generalization of the results of this paper is to apply 
them to dynamical situations leading to different 
vacua, i.e. $G(t<0)\neq G(t>T)$ and $\ket{0,in}\neq\ket{0,out}$. 
As we can see in section 5 and 6, then it is nessecary to distinguish between 
the particle production  
due to the dynamical Casimir effect and the one which results already from the 
comparision of different vacua. 
The investigation of non-perfect conducting boundaries does not seem to be 
feasable in a straightforward manner within the  
canonical formalism presented above. This requires more involved studies. 
 
\subsection{Comparision with other results} 
 
Applying our approach to the dynamical parallel-plate configuration we also 
recover most of the results 
obtained earlier (see \cite{b1}-\cite{p0}) provided the used approximations 
are taken into account carefully.
E.g., for the example investigated in \cite{p9} : 
$G(t)=(0,L_0[1+\varepsilon\sin(2\omega_1 t)])$ for $0<t<T$ 
with $\omega_1=\pi/L_0$ and $\omega_1 T \gg 1$
we reproduce the obtained result
for $\varepsilon\omega_1 T \ll 1$ :

\begin{equation}
N_1=N^S_1=\frac{1}{4}(\varepsilon\omega_1 T)^2
\quad .
\end{equation}  

For the radiation of a single mirror (see \cite{m1}-\cite{m5}) it is possible 
to compare  
the total radiated energy from (\ref{rad}) with the formula (3.15) of Ref. 
\cite{m4}: 
 
\begin{equation} 
\label{ford}
E=(6\pi)^{-1}\int\limits^\infty_{-\infty}\dot V^2\;dx_0     
\end{equation} 
 
with $V=\dot\eta$ and $dx_0=dt$ (The $\ddot V^2$ in the original formula is 
probably a printing error.) 
The result is the same because the additional factor of $2$ in (\ref{ford}) is 
due to the fact,
that (\ref{rad}) in contrast to (\ref{ford}) describes only the particles 
radiated into $G(t)$, 
i.e. the emission to the right. 
It could be interesting to compare the canonical formalism presented in this 
paper with that of Ford, Vilenkin  
(see \cite{m4}), Moore (see \cite{p1}) and that of Fulling, Davies 
(see \cite{m1}, \cite{m2} and also \cite{m3}). 
The main differences are:\\ 
-In our approach the dynamics is governed by the Hamiltonian, 
but otherwise calculated using a coordinate transformation or Green's 
functions.\\ 
-The regularization by means of the point splitting method, usually applied to 
local quantities as for  
deriving the renormalized energy-momentum tensor, is expected to provide 
results that are independent 
 of the infinitesimal displacement vector $\epsilon^\mu$, is avoided  in the 
canonical formalism in favour of 
regularization procedures (if necessary) applied on mode sums.\\ 
Under which conditions both approaches will lead to a unique result requires 
careful investigations for each  
configuration of boundaries under consideration. E.g., one needs to clarify 
 in which coordinate system $\left< \hat T_{\mu\nu} \right> $ should be 
calculated  
and renormalized (see section 6). 
      
\section{Appendix} 
\appendix 
 
We have to introduce a complete set of real and orthonormal  
eigenfunctions of the Laplace operator, satisfying the Dirichlet boundary 
conditions 
$f_\alpha=0$ at $\partial G(t)$:

\begin{equation}  
\label{g2} 
\int\limits_{G(t)}\,dV\,f_{\alpha}\,f_{\beta} 
=\delta(\alpha,\beta) \quad , 
\end{equation} 
 
\begin{equation} 
\label{g3} 
\int\limits_{G(t)}\,dV\, 
(\nabla f_{\alpha})\,(\nabla f_{\beta}) 
=(\Omega_{\alpha}(t))^2\;\delta(\alpha,\beta) \quad , 
\end{equation} 
 
\begin{equation}  
\label{g4} 
\sumint\limits_\alpha\,\, f_{\alpha}(\vec r\,)\,f_{\alpha}(\vec r\,') 
=\delta(\vec r - \vec r\,')
\hspace{2cm} {\rm in} \quad G(t) \quad . 
\end{equation}

For the moving-mirrors-configuration the domain $G(t)$ can be expressed as 
follows: 
 
\begin{equation}  
\label{a1} 
G(t)=I(t) \otimes G^{\perp}  
\end{equation} 
 
with a time-dependent one-dimensional $I(t)$ determining the separation and 
the time-independent subdomain $G^\perp$ 
defining the areas of the plates.\\ 
Then the eigenfunctions $f_\alpha$ factorize: 
 
\begin{equation}\ 
\label{a10} 
f_\alpha(\vec r,t)=f^\|_n(z,t)\;f^\perp_r(\vec r^\perp) 
\end{equation} 
 
with $\alpha=(n,r)$ , $\beta=(m,s)$   and $z$ denotes the parallel component of 
the position vector $\vec r$. Accordingly,  
$\;f^\perp_r=0$ at $\partial G^\perp\;$ and $\;f^\|_n=0$ at $\partial I(t)$ .\\ 
The frequencies decompose into two corresponding parts:  
 
\begin{equation} 
\label{a11} 
\Omega^2_\alpha(t)= 
[\Omega^\perp_r]^2+[\Omega^\|_n(t)]^2 \quad . 
\end{equation} 
 
In $G^\perp$: 
 
\begin{equation}  
\label{a2} 
\int\limits_{G^{\perp}}\,dV^{\perp}\,f^{\perp}_{r}\,f^{\perp}_{s} 
=\delta(r,s) \quad , 
\end{equation} 
 
\begin{equation} 
\label{a3} 
\int\limits_{G^{\perp}}\,dV^{\perp}\, 
(\nabla f^{\perp}_{r})\,(\nabla f^{\perp}_{s}) 
=(\Omega^{\perp}_{r})^2\;\delta(r,s) \quad , 
\end{equation} 
 
\begin{equation}  
\label{a4} 
\sumint\limits_r\,\, f^{\perp}_{r}(\vec a^\perp )\,f^{\perp}_{r}(\vec b^\perp ) 
=\delta(\vec a^\perp - \vec b^\perp)\quad . 
\end{equation} 
 
Note that, if we expand $\Phi$ in eigenfunctions only in $G^\perp$ : 
 
\begin{equation} 
\label{a5} 
\Phi(\vec r,t)=\sumint\limits_r\phi_r(z,t)\;f^\perp_r(\vec r^\perp) \quad , 
\end{equation} 
 
where $\forall r\;\phi_r=0$ at $\partial I(t)$, the $n$-dimensional Lagrangian 
becomes a sum of 
effective 1-dimensional Lagrangians:   
 
\begin{equation} 
\label{a6} 
L=\sumint\limits_r\int\limits_{I(t)}\,dz\frac{1}{2}(\dot\phi^2_r-
(\nabla\phi_r)^2-(\Omega^\perp_r)^2\phi^2_r)\quad . 
\end{equation} 
 
In $I(t)$ we may expand the $\phi_r$ according to: 
 
\begin{eqnarray} 
\phi_r(z,t) = \sumint\limits_{n} \,q_{(n,r)}(t)\;f^\parallel_n(z,t)\quad , 
\end{eqnarray} 
 
that fulfil the relations 
 
\begin{equation}  
\label{a7} 
\int\limits_{I(t)}\,dz\,f^\parallel_n(z,t)\,f^\parallel_m(z,t) 
=\delta(n,m)\quad , 
\end{equation} 
 
\begin{equation}  
\label{a8} 
\int\limits_{I(t)}\,dz\,(\nabla f^\parallel_n(z,t)) 
(\nabla f^\parallel_m(z,t)) 
=(\Omega^{\|}_n(t))^2\; \delta(n,m)\quad , 
\end{equation} 
 
and 
 
\begin{equation} 
\label{a9}  
\sumint\limits_n\,\, f^{\|}_{n}(a,t)\,f^{\parallel}_{n}(b,t) 
=\delta(a-b)
\hspace{2cm}{\rm in}\quad I(t) \quad . 
\end{equation} 
 
For instance, we may specify the parallel-plate configuration via $I(t)=
(\eta(t);l(t)+\eta(t))$ together 
with the eigenmodes refering to the constrained dimension: 
 
\begin{equation} 
\label{a12}  
f^\|_n(z,t)=\sqrt{\frac{2}{l(t)}}\sin\left(\frac{n\pi}{l(t)}[z-\eta(t)]\right) 
\end{equation} 
 
with eigenfrequencies: 
 
\begin{equation} 
\label{a13} 
\Omega^\|_n(t)=\frac{n\pi}{l(t)}\quad . 
\end{equation} 
The coupling matrix then reads  
 
\begin{equation} 
\label{a14}  
{\cal M}_{\alpha\beta}(t)=-{\cal M}_{\beta\alpha}(t)= 
\frac{\dot l (t)}{l(t)}  { \cal G }_{\alpha\beta} 
+\frac{\dot\eta(t)}{l(t)}  { \cal A }_{\alpha\beta}\quad , 
\end{equation}

where for  $n\neq m$ : 
 
\begin{equation} 
\label{a15}  
{\cal A}_{\alpha\beta}= 
[(-1)^{m+n}-1]\frac{2mn}{m^2-n^2}\,\delta(r,s)\quad , 
\end{equation} 
 
\begin{equation} 
\label{a16}  
{\cal G}_{\alpha\beta}=(-1)^{m+n}\frac{2mn}{m^2-n^2}\,\delta(r,s) 
\end{equation} 
 
and with $\;{\cal G}_{\alpha\beta}={\cal A}_{\alpha\beta}=0\;$ for $n=m$.\\ 
Considering one single mirror as the limiting case $G(t)=(\eta (t);\infty)$ it 
follows: 
 
\begin{equation} 
\label{a17}  
{\cal M}_{\alpha\beta}(t)= 
\dot\eta (t)\frac{2}{\pi}{\cal P}
\left(\frac{\Omega_\alpha\Omega_\beta}{\Omega^2_\alpha-\Omega^2_\beta}\right)
\quad , 
\end{equation} 
 
where ${\cal P}$ denotes the principal value. Special care is required in any  
calculation involving such distribution-like functions (e.g. order of 
integration); 
for instance to obtain the following result: 
 
\begin{equation}  
\label{end} 
{\cal M}_{\alpha\gamma}(t){\cal M}_{\beta\gamma}(t)= 
\dot\eta^2(t)\;\Omega^2_\alpha\;\delta(\alpha,\beta)\quad . 
\end{equation}

\end{document}